\documentclass[a4paper,english,preprint]{revtex4}
\usepackage{times}
\usepackage[T1]{fontenc}
\usepackage[latin1]{inputenc}
\usepackage{array}
\usepackage{color}
\usepackage{graphicx}

\makeatletter

\providecommand{\tabularnewline}{\\}

\usepackage{pslatex}

\usepackage{babel}
\makeatother
\begin{document}

\title{\noindent Free energy calculations of elemental sulphur crystals
\emph{via} molecular dynamics simulations }

\author{C. Pastorino and Z. Gamba}

\affiliation{\emph{Department of Physics, Comisión Nacional de Energía Atómica}
\emph{CAC, Av. Libertador 8250, (1429) Buenos Aires, Argentina.} }

\email{clopasto@cnea.gov.ar, gamba@cnea.gov.ar.}

\begin{abstract}
Free energy calculations of two crystalline phases of the molecular
compound $S_{8}$ were performed \emph{via} molecular dynamics simulations
of these crystals. The elemental sulphur $S_{8}$ molecule model used
in our MD calculations consists of a semi-flexible closed chain, with
fixed bond lengths and intra-molecular potentials for its bending
and torsional angles. The intermolecular potential is of the atom-atom
Lennard-Jones type. Two free energy calculation methods were implemented:
the accurate thermodynamic integration method proposed by Frenkel
and Ladd \cite{free-frenkel-ladd1} and an estimation that takes into
account the contribution of the zero point energy and the entropy
of the crystalline vibrational modes to the free energy of the crystal.
The last estimation has the enormous advantage of being easily obtained
from a \emph{single} MD simulation. Here we compare both free energy
calculation methods and analyze the reliability of the fast estimation
\emph{via} the vibrational density of states obtained from constrained
MD simulations. New results on $\alpha$- and $\alpha$'-$S_{8}$
crystals are discussed.
\end{abstract}

\date{\today}

\maketitle

\section{Introduction}

In recent papers we studied the crystalline phases of the molecular
compound $S_{8}$, the most abundant natural form of elemental sulphur
around ambient pressure and temperature (STP). Three crystalline allotropes
of this compound are known: $\alpha$-$S_{8}$ phase is orthorhombic
with 16 molecules in the non-primitive unit cell \cite{meyer1,steudel}.
If several samples of this crystal are slowly heated, some of them
will show a phase transition to $\beta$-$S_{8}$ at 369 K, which
in turn melts at 393 K \cite{steudel}, but most of them show a metastable
melting point at 385.5 K \cite{meyer1,steudel}. Monoclinic $\beta$-$S_{8}$
is usually obtained from the melt \cite{steudel} and has 6 molecules
per primitive unit cell, two of them orientationally disordered above
198 K \cite{beta1}. Monoclinic $\gamma$-$S_{8}$ has four molecules
in a pseudo-hexagonal close packed unit cell, with a density 5.8\%
higher than $\alpha$-$S_{8}$ at STP \cite{meyer1}.

Using a very simple molecular model consisting of a cyclic semi-flexible
chain of 8 atoms, with constant S-S bond lengths, it is possible to
reproduce many features of the complex phase diagram of the $S_{8}$
molecule \cite{pap1,pap2}. The intra-molecular potential model includes
a harmonic bending potential for S-S-S angles and a double well for
the torsional angles \cite{cardini1}, the intermolecular potential
used was a simple \emph{Lennard-Jones} non-bonded atom-atom interaction,
the details and potential parameters are given in the Section \ref{sec:Intra--and-intermolecular}.
Using this simple interaction model we could reproduce, via a series
of classical constant temperature- constant pressure simulations \cite{pap1,pap2},
the following experimental facts: the crystalline structure, the configurational
energy and the lattice, bending and torsional dynamics (as given by
the calculated density of vibrational modes) of $\alpha$-, $\beta$-
and $\gamma$-$S_{8}$ for T$\geq$200 K; the orientational dynamical
order - disorder phase transition of $\beta$-$S_{8}$ and, finally,
the solid-liquid phase transition of a cubic disordered sample was
calculated near the experimental value \cite{pap1,pap2}. 

\begin{figure}[!htb]
\includegraphics[%
  scale=0.5]{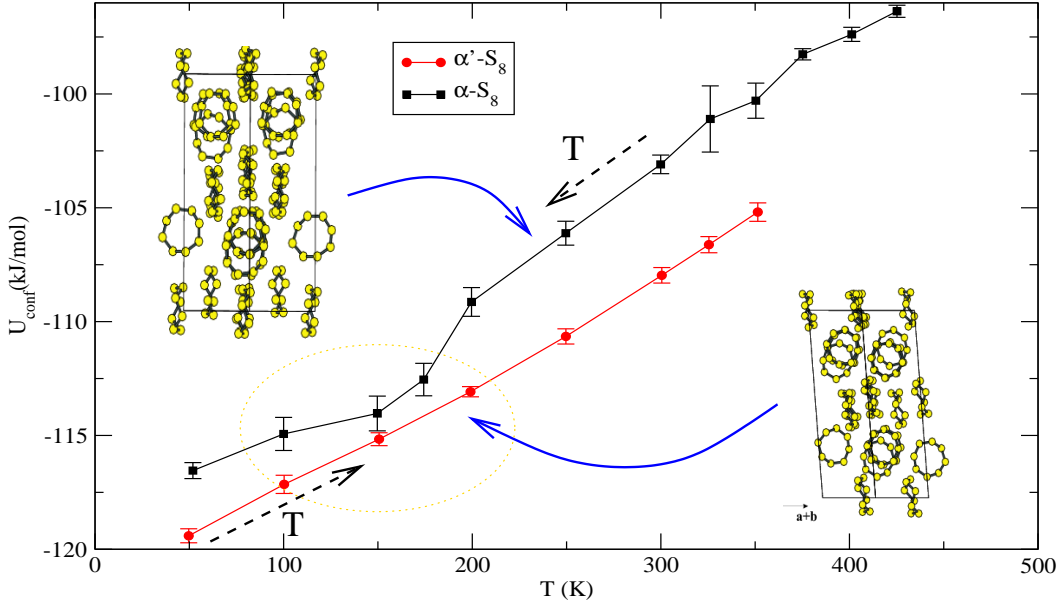}

\caption{Configurational energy of $\alpha$ and $\alpha$'-$S_{8}$ as obtained
in ref. \cite{pap1}. The structural transition to $\alpha$'-$S_{8}$
is observed spontaneously below $200\, K$. \label{fig:econf}}
\end{figure}

Nevertheless, we found a fact that cannot be reproduced by this simple
molecular model \cite{pap1,pap2}: when the temperature of a $\alpha$-$S_{8}$
MD sample of 288 molecules is lowered below 200 $K$, our orthorhombic
$\alpha$-$S_{8}$ sample shows a structural phase transition to a
monoclinic phase, with a molecular array similar to that of $\alpha$-$S_{8}$
and that we called $\alpha'$-$S_{8}$ \cite{pap2}. This has not
been experimentally observed. This spontaneous change was most probably
due to the large fluctuations associated to a relatively small sample.
A larger sample of 512 molecules didn't show this distortion, but
its configurational energy and volume per molecule remained, nevertheless,
higher than those calculated for $\alpha'$-$S_{8}$ \cite{pap2}.
Fig. \ref{fig:econf} shows both structures and the configurational
energy vs. temperature, that of $\alpha$-$S_{8}$ was calculated
by slowly decreasing T in steps of 25 to 50 K. The same study for
$\alpha'$-$S_{8}$ is performed by increasing T, after a long stabilization
and annealing to obtain a totally ordered structure at the starting
point of 50 K, this is the reason for the different $U_{conf}$ values,
in fig. \ref{fig:econf}, at this point. Experimental measurements
of the $\alpha$-$S_{8}$ structure at 300 and 100 K clearly disregard
a structural phase transition, unless a metastable state has been
experimentally measured \cite{alfa2,exp-alemanes}, and this fact
shows the limits of applicability of the simple molecular model.

In refs. \cite{pap2,pap3} we analyzed if an anisotropic non- bonded
atom-atom intermolecular potential model was able to reproduce all
the experimental facts. Using the program GAMESS \cite{gamess}, we
performed \emph{ab initio} calculations of the electronic density
distribution of the $S_{8}$ molecule, and the measured anisotropy
of this distribution around each atom due to the atomic lone pairs.
The calculated electronic density reproduces the experimental crystalline
measurements of Coppens \emph{et. al.} \cite{alfa3}, with a lone
pair center at 0.7\AA ~of the S location; nevertheless, at the nearest
crystalline distances found between non-bonded atoms, the atomic anisotropy
is low. The MD simulations performed with this anisotropic potential
did not improve the previous result (shown in Fig. \ref{fig:econf}),
unless a quite unrealistic atomic anisotropy is used \cite{pap3,pap4}. 

It has to be pointed out that the search of a reliable molecular model
for elemental sulphur molecules is extremelly useful due to the practical
impossibility of performing quantum mechanical simulations of the
complex phase diagram of these molecular crystals, with a large number
of atoms in the primitive cells.

In this paper we decided to implement a reliable calculation of the
free energy of $\alpha$-$S_{8}$ and $\alpha'$-$S_{8}$ crystals
in order to review the conclusions obtained with the first molecular
model \cite{pap1,pap2}, based on an estimation of the free energy
that takes into account the contribution of the vibrational modes
to the zero point energy and entropy of the sample \cite{pap1,pap2}.
Our present calculations are useful to check our previous free energy
estimations and, at the same time, for the comparison of both free
energy calculation methods. The usefulness of this comparison is due
to the fast way in which the free energy can be estimated, within
the quasi-harmonic approximation, using the data of a single MD simulation.

In the following sections we give the details of the implemented free
energy calculations, the inter- and intra-molecular potential model
used, the performed MD simulations, the obtained results and our conclusions. 

\section{Free energy calculations}

In order to obtain reliable free energy values we implemented the
thermodynamic integration method of Frenkel and Ladd \cite{free-frenkel-ladd1},
including the two correction terms given in refs. \cite{free-frenkel-correction,understanding}.
From the results of a single MD simulation, we also estimated the
free energy in the quasi-harmonic approximation, that is usually employed
in lattice dynamics calculations \cite{bornhuang}. In Section \ref{sec:md}
we give the details of our performed MD simulations.

\subsection{Thermodynamic integration method}

This method proposes a thermodynamic integration along a reversible
path between the system we are interested in and another one for which
the free energy can be analytically calculated \cite{free-frenkel-ladd1,free-frenkel-correction,understanding}.
For crystalline samples the reference system is an Einstein crystal,
for which the internal energy is calculated as:

\[
U_{Einstein}=U(\{\mathbf{r}_{0}^{N}\})+\sum_{i=1}^{N}\alpha(\mathbf{r}_{i}-\mathbf{r}_{0})^{2}\,,\]
where $\{\mathbf{r}_{0}^{N}\}$ is the set of coordinates of the minimum
energy configuration (the equilibrium atomic coordinates) and $U(\{\mathbf{r}_{0}^{N}\})$
its potential energy, the second term is the harmonic potential energy
of the N oscillators. The force constant $\alpha$ is taken equal
for all degrees of freedom and its value, $\alpha=$18.0844 kJ/mol,
is chosen so as to obtain a mean square displacement equal to the
average one in the real crystal{\small : $\left\langle \Delta r^{2}\right\rangle _{Einstein}\sim\left\langle \Delta r^{2}\right\rangle _{real}$}.
The potential energy difference between the real and the reference
crystals can be calculated along a reversible artificial pathway that
links both systems. At any intermediate point of the path, given by
the coordinate $\lambda$ $(0\leq\lambda\leq1)$, the local value
of $U$ is

\begin{equation}
\widetilde{U}(\lambda)=U(\{\mathbf{r}_{0}^{N}\})+(1-\lambda)[U(\{\mathbf{r}^{N}\})-U(\{\mathbf{r}_{0}^{N}\})]+\lambda\sum_{i=1}^{N}\alpha(\mathbf{r}_{i}-\mathbf{r}_{0})^{2}\,.\label{eq:ti1}\end{equation}

So, for $\lambda=0$ the potential energy of the real crystal is recovered
and, for $\lambda\textrm{=1}$ the potential energy is that of the
Einstein crystal:

\[
\widetilde{U}(\lambda=0)=U_{sist}\]
and \[
\widetilde{U}(\lambda=1)=U_{Einstein}\,.\]

The free energy difference between the real and the reference crystal
is given by \cite{free-frenkel-ladd1,understanding}:

\[
F(\lambda=0)-F(\lambda=1)=-\int_{0}^{1}d\lambda(\frac{\partial F(\lambda)}{\partial\lambda})=-\int_{0}^{1}d\lambda\langle\frac{\partial\widetilde{U}(\lambda)}{\partial\lambda}\rangle_{\lambda},\]

and deriving eq. \ref{eq:ti1}:

\begin{equation}
F=F_{Einstein}+\int_{\lambda=1}^{\lambda=0}d\lambda\left\langle \sum_{i=1}^{N}\alpha(\mathbf{r}_{i}-\mathbf{r}_{0})^{2}-[U(\mathbf{r}^{N}{})-U(\mathbf{r}_{0}^{N}{})]\right\rangle _{\lambda}\,.\label{eq:ti2}\end{equation}
The free energy of the Einstein \emph{}crystal, with a fixed center
of mass for the MD sample, is \cite{polson-frenkel1}: \emph{}

\begin{equation}
\frac{F_{Einstein}}{N}=U(\mathbf{r}_{0}^{N})-\frac{3(N-1)}{2N\beta}\ln(\frac{2\pi}{\alpha\beta})-\frac{3\ln N}{2N}-\frac{\ln V_{0}}{\beta N}+3\ln(\Lambda)\,,\label{eq:ti3}\end{equation}

where $\Lambda=\frac{h}{\sqrt{2\pi mk_{B}T}}$ is the de Broglie thermal
wavelength , $\beta=\frac{1}{k_{B}T}$, $N$ is the number of atoms
and $V_{0}$ is the equilibrium volume of the sample. 

If $N_{mol}$ is the number of molecules of the system, the final
expression, including a correction term due to the center of mass
constraint of the sample \cite{polson-frenkel1,free-frenkel-ladd1,understanding}
and a second correction term due to its finite size \cite{free-frenkel-correction},
is:

\begin{equation}
\frac{\beta F}{N_{mol}}=3\frac{N}{N_{mol}}\ln(\Lambda)-\frac{3(N-1)}{2N_{mol}}\ln(\frac{2\pi}{\beta\alpha})-\frac{3\ln(N)}{2N_{mol}}-\frac{\ln V_{0}}{N_{mol}}-\frac{\beta}{N_{mol}}\int_{0}^{1}\left\langle U_{ein}-U\right\rangle _{\lambda}\,.\label{eq:ti4}\end{equation}

The accuracy of this calculation is determined by a correct evaluation
of the last term in eq. \ref{eq:ti4}, where long MD simulations are
needed to measure the integrand for each value of $\lambda$. 

For all values of $\lambda$, the center of mass of the sample is
held fixed at the origin because the {}`springs' of the reference
Einstein crystal are tied to the equilibrium atomic positions. This
is important to avoid divergences of the integrand in eq. \ref{eq:ti4}
when $\lambda$ is close to 0 \cite{free-frenkel-ladd1,understanding}. 

The third term in eqs. \ref{eq:ti3} and \ref{eq:ti4} includes the
change of the sign mentioned in ref. \cite{free-frenkel-correction},
with respect to the original work of Frenkel and Ladd \cite{free-frenkel-ladd1},
due to a new consistent calculation of the partition function of the
constrained CM system \cite{free-frenkel-correction}. This term is
negligible in a large system but must be taken into account to compute
absolute free energies of a typical MD sample.

\subsection{Quasi-harmonic approximation \label{sub:Quasi-harmonic-approximation}}

Following Born and Huang \cite{bornhuang}, the free energy of a harmonic
crystal with $N_{d}$ degrees of freedom ( i. e. $N_{d}$ independent
harmonic oscillators) can be calculated as:

\begin{equation}
F=\sum F_{i}=U_{conf}+\frac{1}{2}\sum_{i=1}^{N_{d}}h\nu_{i}+k_{B}T\sum_{i=1}^{N_{d}}\,\ln(1-e^{-\frac{h\nu_{i}}{k_{B}T}})\,\textit{,}\label{eq:f-con-vds}\end{equation}

where $U_{conf}$ is the potential energy and the $\nu_{i}$ are the
vibrational frequencies of a system with $N_{d}$ degrees of freedom.

We have implemented this method, as in ref. \cite{pap1,pap2}, in
two steps. First we calculate the atomic velocities self correlation
function $C(t)$, using the MD data stored in a free trajectory in
the phase space of the sample:

\[
C(t)=\frac{\left\langle v(t).v(0)\right\rangle }{\left\langle v(0).v(0)\right\rangle }\,\textit{,}\]
 where $<>$ imply averages over atoms and over different initial
times.

In a second step we calculate the vibrational density of states $D(\nu)$
\emph{via} a Fourier transform of $C(t)$, and replace the sums over
frequencies in eq. \ref{eq:f-con-vds} by an integral on frequencies,
weighted by the calculated density of vibrational modes. We have to
take into account that the normalization factor for $D(\nu)$ is such
that $\int\, D(\nu)\, d\nu=N_{d}.$ In our case is $N_{d}=16N_{mol}$,
because we have 8 bond constraints per molecule. 

We can then calculate the internal energy, the entropy and free energy
as \cite{bornhuang}:

\begin{equation}
F=U_{conf}+\frac{1}{2}\int d\nu\, D(\nu)\, h\nu+k_{B}T\int d\nu\, D(\nu)\,\ln(1-e^{-\frac{h\nu}{k_{B}T}})\,\,,\label{eq:free-vds}\end{equation}

\[
S=-(\frac{\partial F}{\partial T})_{V}=\int\frac{e^{-\frac{h\nu}{k_{B}T}}{}}{1-e^{-\frac{h\nu}{k_{B}T}}}\frac{h\nu}{T}D(\nu)d\nu\,\,,\]

\[
U=F-TS=U_{conf}+\frac{1}{2}\int d\nu\, D(\nu)\, h\nu+\int\frac{e^{-\frac{h\nu}{k_{B}T}}{}}{1-e^{-\frac{h\nu}{k_{B}T}}}\frac{h\nu}{T}D(\nu)d\nu\,.\]

The approximations involved in our quasi-harmonic calculations, with
the results of a single MD simulation, are the following:

\begin{itemize}
\item $U_{conf}$ is calculated by performing averages, over time and molecules,
of the potential energy per molecule calculated in the MD run, $U_{conf}=<U>$,
it is not the value of the minimum potential energy $U_{0}$ of the
Einstein crystal in equation \ref{eq:ti3}.
\item The calculated vibrational density of states $D(\nu)$ of the MD crystalline
sample, due to its finite size ( a few number of primitive unit cells
are included in it) does not give an accurate measurement of the density
given by the real dispersion curves, the frequencies are only measured
in a few points of the reciprocal space.
\item The anharmonic frequencies of the sample can be accurately measured
with a MD simulation, but eq. \ref{eq:free-vds} is called the 'quasi-harmonic'
approximation because eq. \ref{eq:f-con-vds} is exact only for harmonic
potentials.
\end{itemize}

\section{Intra- and intermolecular potential model\label{sec:Intra--and-intermolecular}}

The potential model is that of refs. \cite{pap1,pap2}. The flexible
molecular model includes all low frequency molecular distortions that
mix with lattice modes and can therefore be of relevance in a possible
structural phase transition. The S-S bond distances are kept constant
($d_{SS}=2.0601$ \AA) because the stretching modes are well above
in energies ($\nu>400\, cm^{-1}$) than the rest of the vibrational
modes ($\nu<250cm^{-1}$). 

We must stress here that bond constraints can have nonnegligible contributions
to the free energy calculations. Such contributions have been studied
elsewhere , with different methods and techniques \cite{constraints-estima1,constraint-estima2,constraints-estima3}.
The main conclusions of those explicit calculations, using both MD
\cite{constraint-estima2} and MCTI \cite{constraints-estima1}, are
that the influence of bond constraints is actually negligible when
the bond length is not changed from the initial to the final state.
This is the case in our $S_{8}$ phases, in which the primitive cell
will change and the molecules will distort from one phase to the other
but the molecules bond length is held fixed within the same value
for every calculation presented in this work. If changes in bond length
between initial and final states of the free energy calculation were
involved, an explicit constraint contribution to the free energy must
be considered as explained in refs. \cite{constraints-estima1,constraints-estima3}.
For the sake of completeness, a simple estimation of the change in
the probability density between the standard $NVT$ and the {}`bond-constrained'
ensemble of $S_{8}$ is given in the Appendix.

The bending intramolecular potential for S-S-S angles is harmonic
\cite{cardini1},

\[
V(\beta)=\frac{1}{2}C_{\beta}(\beta-\beta_{0})^{2}\]

with a force constant of C$_{\beta}$=25725 $k_{B}/$rad$^{2}$ and
$\beta_{0}$= 108 deg. The intramolecular potential for torsion angles
is a double well \cite{cardini1},

\[
V(\tau)=A_{\tau}+B_{\tau}\cos(\tau)+C_{\tau}\cos^{2}(\tau)+D_{\tau}\cos^{3}(\tau)\,\textit{,}\]

with $A_{\tau}$=57.192 k$_{B}$, $B_{\tau}$=738.415 k$_{B}$, $C_{\tau}$=2297.880
k$_{B}$ and $D_{\tau}$=557.255 k$_{B}$. These parameters describe
a double well with minima at $\tau$=180 ${_{-}^{+}}$ 98.8 deg.,
and a barrier height of about 9 kJ/mol at $\tau$=180 deg. The barrier
height at $\tau$=0 deg. is of 30 kJ/mol, out of the range of energies
explored in these simulations.

The intermolecular potential is of the non-bonded atom-atom Lennard-Jones
type, with parameters $\varepsilon$=1.70 kJ/mol and $\sigma$=3.39~\AA\ 
\cite{cardini1,pap1}. The cut-off radius of the interactions is 15~ \AA\ 
and correction terms, to the energy and pressure, due to this finite
cut-off are taken into account by integrating the contribution of
an uniform distribution of atoms for distances larger than our cut-off
of 15~\AA.

\section{Molecular dynamics simulations\label{sec:md}}

The MD simulations and samples are entirely identical to those previously
performed in refs. \cite{pap1,pap2}. The bond constraints are held
fixed with the SHAKE algorithm \cite{tildesley} and the temperature
is maintained using the Nosé-Hoover chains method \cite{nose-hoover-chains},
although the same behavior was found using the standard Nosé method
\cite{nose-hoover1,nose-hoover2} in our NPT ensemble simulations.

Taking the equilibrated samples of $\alpha$- and $\alpha$'-S$_{\textrm{8}}$
from our simulations in refs. \cite{pap1,pap2} as starting point,
we first did a careful measurement of their crystalline structures
(i. e. the lattice parameters and volume $V_{0}$) by averaging the
calculated values over long runs performed in the NPT ensemble. The
time step of the simulations was of 0.01 ps, the thermalization of
the samples was of 80000 time steps (800 ps) and they were measured
in the following 30000 (300 ps) time steps. The lattice parameters
of each one of the four samples, together with their averaged atomic
positions, were taken as the equilibrium atomic locations for the
corresponding Einstein crystals.

The next series of MD simulations, as a function of the $\lambda$
parameter $(0<\lambda<1)$, were performed at constant volume, with
a standard time step of 0.005 ps and runs of 160000 time steps (800
ps) of thermalization followed by a 40000 time steps (200 ps) of a
free trajectory in the phase space, that were used to measure the
systems. 

For $\lambda$ values close to 1, we had to reduce the time step up
to a minimum of 0.0001 ps in order to obtain the same accuracy in
the total energy as for the other values of $\lambda$. Accordingly,
the total number of time steps were increased so as to accumulate
nearly the same total time of thermalization and storage of the other
cases. 

As in refs. \cite{polson-frenkel1,polson-frenkel2}, ten values of
the parameter $\lambda$ are defined by a ten point Gauss-Legendre
quadrature method, used to resolve the thermodynamic integration on
the $\lambda$ coordinate (last term en eq. \ref{eq:ti4}).

\section{Results}

\subsection{Thermodynamic integration method}

\begin{figure}[!htb]
\includegraphics[%
  scale=0.5]{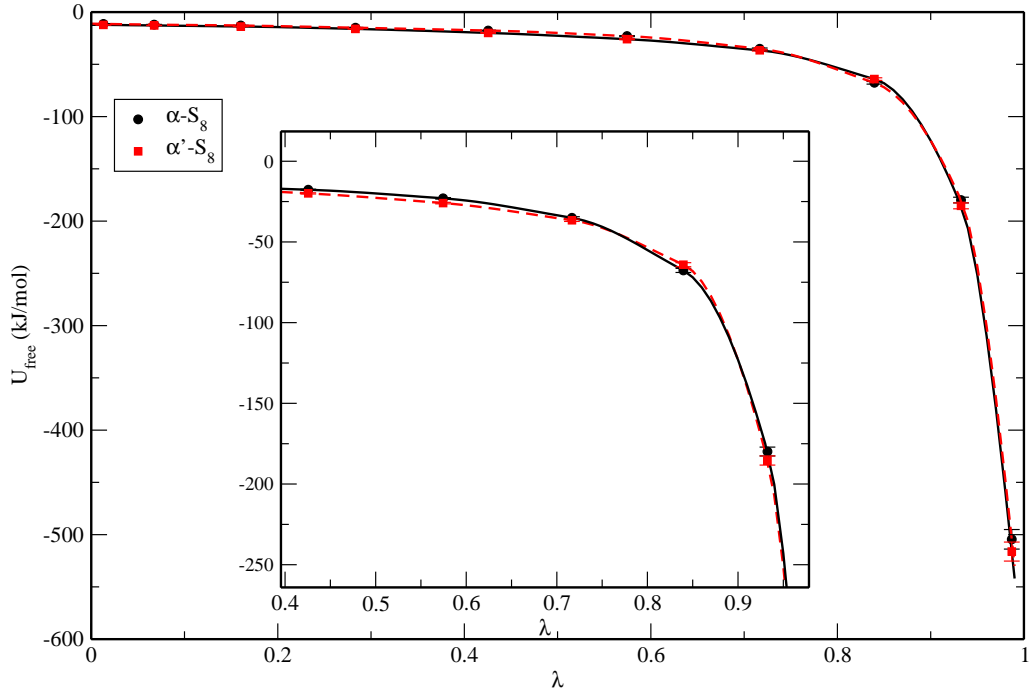}

\caption{$\left\langle U_{free}\right\rangle $ for $\alpha-$ and $\alpha'$-$S_{8}$
at 300K. Each point is a MD run. This curve is the integrand of the
last term in equation \ref{eq:ti4}, from which the free energy is
obtained using the Frenkel Ladd thermodynamic integration scheme \label{fig-u-free}.
The lines are a guide to the eyes and were calculated with the Akima
interpolation algorithm.}
\end{figure}

Fig. \ref{fig-u-free} shows the values of $U_{free}(\lambda)=\left\langle U_{Einstein}-U\right\rangle _{\lambda}$,
obtained with averages over the lasts 200 ps of each run, for $\alpha$-
and $\alpha$'-$S_{8}$ at 300 K. Similar curves $U_{free}(\lambda)$
are calculated for $\alpha-$ and $\alpha'$-$S_{8}$ at 100K, except
that for $\lambda\sim$1 $<U_{free}>$ is about $-260$ kJ/mol. For
both temperatures the values for the $\alpha'$-$S_{8}$ sample are
systematically lower than those of $\alpha$-$S_{8}$, the difference
increases for increasing values of $\lambda$. 

Note that the values $U_{free}(\lambda)$ near $\lambda\sim$1 are
those that weight more in the calculation of the integral in eq. \ref{eq:ti4}.
$\lambda\sim$1 implies strong springs of Einstein crystals and weak
interactions of the original system, in this case the slope of $<U_{free}(\lambda\sim1)>$
has a large increment because each atom is subjected almost only to
the harmonic force of the Einstein crystal, allowing molecular configurations
very different from those of the real system. Each atom shows a mean
square displacement that depends only on temperature and not in the
forces of the real system. At higher temperatures the interactions
are more repulsive (because of the shorter mean interatomic distances)
and $<U_{free}>$ takes more negative values ($U$ has a negative
sign in the integrand of eq. \ref{eq:ti4}). This can be verified
noting that for $T=100$ K the minimum $<U_{free}>$ is approximately
$-260$ kJ/mol, almost half the value at $300$ K (-510 kJ/mol, see
figure \ref{fig-u-free}).\\

The contributions of the different terms to the total free energy
of eq. \ref{eq:ti4}, in adimensional units, are included in Table
\ref{tab-detalle-ti}. The first three rows present the same values
for all the cases, the first two of them are large terms that almost
cancel each other. The next three rows show the terms that are calculated
different in each case, and are obtained from the MD simulation data.
The main contributions are from the $U_{0}$ static term and from
the thermodynamic integration term. In the last row the adimensional
free energy values for all the cases are shown. 

\begin{table}[!htb]
\begin{tabular}{|c|c|c|c|c|}
\hline 
Phase/Temp.&
\textbf{\small $\alpha$-$S_{8}$ ($T=100K$)}&
\textbf{\small $\alpha'$-$S_{8}$($T=100K$)}&
\textbf{\small $\alpha$-$S_{8}$ ($T=300K$)}&
\textbf{\small $\alpha'$-$S_{8}$ ($T=300K$)}\tabularnewline
\hline
\hline 
{\small $3\frac{N}{N_{mol}}\ln\Lambda^{*}$}&
-1162.78&
-1162.78&
-1162.78&
-1162.78\tabularnewline
\hline 
{\small $-\frac{3(N-1)}{2N_{mol}}\ln(\frac{2\pi}{\beta\alpha^{*}})$}&
1148.94&
1148.94&
1148.94&
1148.94\tabularnewline
\hline 
{\small -$\frac{3\ln(N)}{2N_{mol}}$}&
{\small -0.04032}&
{\small -0.04032}&
{\small -0.04032}&
{\small -0.04032}\tabularnewline
\hline
\hline 
{\small $\beta U_{0}$}&
{\small -143.05}&
{\small -146.08}&
{\small -46.91}&
{\small -48.47}\tabularnewline
\hline 
{\small $-\frac{\ln(V_{0}^{*})}{N_{mol}}$}&
-0.02529&
-0.02529&
-0.02540&
-0.02535\tabularnewline
\hline 
{\small $-\frac{\beta}{N_{mol}}\int_{0}^{1}d\lambda\left\langle U_{ein}-U\right\rangle _{\lambda}$}&
{\small 25.1(4)}&
{\small 25.5(4)}&
{\small 21.3(4)}&
{\small 22.0(4)}\tabularnewline
\hline
\hline 
$\frac{\beta F_{einstein}}{N_{mol}}$&
{\small -143.79}&
{\small -146.63}&
{\small -47.65}&
{\small -49.22}\tabularnewline
\hline 
{\small $\frac{\beta F}{N_{mol}}$}&
{\small -131.41}&
{\small -133.97}&
{\small -39.05}&
{\small -39.91}\tabularnewline
\hline
\end{tabular}

\caption{Contribution of the different terms to the free energy value, calculated
by the thermodynamic integration method for both temperatures and
phases. For the sake of comparison each term has been adimensionalized
using the definitions $\Lambda^{*}=\Lambda/\sigma$, $\alpha^{*}=\alpha\sigma^{2}$
and $V_{0}^{*}=V_{0}/\sigma^{3}$ when needed.\label{tab-detalle-ti}}
\end{table}

The integral for thermodynamic integration in eq. \ref{eq:ti4} was
also calculated using numerical integration for the Akima interpolation
curve (see figure \ref{fig-u-free}) and the difference with the Gauss-Legendre
quadrature is within 6.4 \%. The error of this term, included in table
\ref{tab-detalle-ti} was obtained with the same Gauss-Legendre quadrature
using the values $U_{free}(\lambda)+\Delta U(\lambda)$ and $U_{free}(\lambda)-\Delta U(\lambda)$,
where $\Delta U$ is the statistical error of each MD simulation. 

There is a further comment about this free energy calculation method
and its use in systems with multiple constraints, as are the very
commonly used semiflexible molecular models with bond length constraints.
In our simulations the constraints were not only applied to the real
systems but also to the corresponding Einstein crystals. In this way
we computed free energy differences between two constrained systems
(the real and the reference) for each one of the four studied cases.
This is irrelevant as long as we compare the free energy between the
$\alpha$- and $\alpha$'-$S_{8}$ crystals or between them and their
reference systems. However, when computing the absolute free energy
of each crystal, it has to be taken into account that the bond length
constraints affect its MD trajectory in the phase space . The correction
terms for systems with multiple types of constraints were recently
discussed by Otter and Briels \cite{briels-fe-const}, they found
that the bond constraint contribution is not negligible only when
comparing systems of different bond lengths \cite{constraint-estima2,constraints-estima1,constraints-estima3}.

\subsection{Quasi-harmonic approximation }

The absolute values of the free energies of the four samples were
also calculated \emph{via} the contribution of the density of vibrational
modes $D(\nu)$ to the entropy and zero point energy of the sample.
With the measurement of an unique run in the $NPT$ ensemble, for
each one of the four real systems, we obtained the values included
in Table \ref{tab-detalle-vds}. The samples are equilibrated and
measured at a pressure of $0._{-}^{+}0.03$ kbar and the $pV$ term
is not included to allow a direct comparison with the thermodynamic
integration method. The volume per molecule $(V_{m})$, configurational
energy ($U_{conf}$), zero point energy ($E_{Q_{0}}$), entropy ($S$)
and free energy ($F$) are shown. The free energy calculated with
this approximation for various temperatures between 100 and 300 K,
not included here, showed a systematically lower values for $\alpha$'-$S_{8}$.

\begin{table}[!htb]
\begin{tabular}{|c|c|c|c|c|}
\hline 
Phase&
\textbf{\small $\alpha$-$S_{8}$ ($T=100K$)}&
\textbf{\small $\alpha'$-$S_{8}$($T=100K$)}&
\textbf{\small $\alpha$-$S_{8}$ ($T=300K$)}&
\textbf{\small $\alpha'$-$S_{8}$ ($T=300K$)}\tabularnewline
\hline
\hline 
$\left\langle V_{m}\right\rangle $(\AA$^{3}$)&
196.8(3)&
196.0(3)&
203.1(4)&
201.2(2)\tabularnewline
\hline 
$\left\langle U_{conf}\right\rangle \textrm{(kJ/mol)}$&
-114.6(1)&
-116.7(1)&
-105.5(3)&
-108.96(3)\tabularnewline
\hline 
$E_{Q_{o}}$(kJ/mol)&
16.083&
16.285&
15.950&
15.836\tabularnewline
\hline 
$TS$(kJ/mol)&
4.3929&
4.2900&
26.6491&
26.9034\tabularnewline
\hline 
$F$(kJ/mol)&
-102.4(3)&
-103.9(3)&
-123.2(1.3)&
-126.0(1.2)\tabularnewline
\hline
\end{tabular}

\caption{Different quantities involved in the free energy calculations \emph{via}
the density of vibrational modes. $E_{Q_{o}}$stands for the zero
point energy term.\label{tab-detalle-vds}}
\end{table}

\subsection{Comparison of both methods}

Table \ref{tab-comp-free-absoluta} includes the calculated absolute
values of free energy, in kJ/mol units, for $\alpha-$ and $\alpha'-S_{8}$
at 100 and 300 K, using both methods. 

We expect higher accuracy in the absolute free energy values for the
thermodynamic integration method, mainly at higher temperatures when
the harmonic approximation is worst than in the case of 100 K, due
to the fact that anharmonic effects are expected to increase at higher
temperatures.

\begin{table}[!htb]
{\small }\begin{tabular}{cc}
{\small }\begin{tabular}{|c|c||c|}
\hline 
\textbf{\textcolor{black}{\small $\mathbf{T=}1\mathbf{00K}$}}&
{\small $\mathbf{F}_{TI}\mathbf{(kJ/mol)}$}&
$\mathbf{F}_{QH}\mathbf{(kJ/mol)}$\tabularnewline
\hline
\hline 
\textcolor{black}{\small $\alpha$-$S_{8}$}&
{\small -109.3(4)}&
-102.4(3)\tabularnewline
\hline 
\textcolor{black}{\small $\alpha'$-$S_{8}$}&
{\small -111.(3)}&
-103.9(3)\tabularnewline
\hline
\end{tabular}&
{\small }\begin{tabular}{|c|c||c|}
\hline 
\textbf{\textcolor{black}{\small $\mathbf{T=300K}$}}&
{\small $\mathbf{F}_{TI}\mathbf{(kJ/mol)}$}&
$\mathbf{F}_{QH}\mathbf{(kJ/mol)}$\tabularnewline
\hline
\hline 
\textcolor{black}{\small $\alpha$-$S_{8}$}&
{\small -130.0(1.0)}&
-123.2(1.3)\tabularnewline
\hline 
\textcolor{black}{\small $\alpha'$-$S_{8}$}&
{\small -132.4(1.0)}&
-126.0(1.2)\tabularnewline
\hline
\end{tabular}\tabularnewline
\end{tabular}{\small \par}

\caption{Comparison of the absolute free energy values using thermodynamic
integration (TI) and Quasi-harmonic approximation (QH). \label{tab-comp-free-absoluta}}
\end{table}

With the thermodynamic integration method the difference of free energies
is calculated about \textcolor{black}{$\Delta F_{\alpha-\alpha'}(T=300K)\sim2\,\frac{kJ}{mol}$}
, being $\alpha$'-$S_{8}$ the more stable phase. Although the absolute
free energy values are not similar, the difference between both phases
is also of about 2 kJ/mol for the quasi-harmonic method. This difference
can be experimentally measured and closely follows that between the
corresponding $U_{0}$ and $U_{conf}$ included in table \ref{tab-comparacion-metod}. 

\begin{table}[!htb]
{\small }\begin{tabular}{|c|c|c|c||c|c|}
\hline 
Phase&
$T(K)$&
$\mathbf{F}_{TI}\mathbf{(kJ/mol)}$&
$\mathbf{F}_{QH}\mathbf{(kJ/mol)}$&
$U_{0}(kJ/mol)$&
$U_{conf}(kJ/mol)$\tabularnewline
\hline
\hline 
{\small $\alpha$-$S_{8}$}&
100.3(1.9)&
{\small -109.3(4)}&
-102.4(3)&
-118.936&
-114.6(1)\tabularnewline
\hline 
{\small $\alpha'$-$S_{8}$}&
100.8(1.9)&
{\small -111.39(34)}&
-103.94(25)&
-121.461&
-116.7(1)\tabularnewline
\hline
{\small $\alpha$-$S_{8}$}&
299.6(5.1)&
{\small -130.27(96)}&
-123.2(1.3)&
-117.008&
-105.5(3)\tabularnewline
\hline
{\small $\alpha'$-$S_{8}$}&
300.9(4.9)&
{\small -132.4(1.0)}&
-126.0(1.2)&
-120.901&
-108.0(3)\tabularnewline
\hline
\end{tabular}{\small \par}

\caption{Comparison of the free energy values for the thermodynamic integration
and the quasi-harmonic approximation methods. The values of the energy
for the mean positions ($U_{o}$) and the mean configuration energy
($U_{conf}$) are also shown for the two phases and the two studied
temperatures.\label{tab-comparacion-metod}}
\end{table}

\section{Conclusions}

In this paper we calculated the free energy of orthorhombic $\alpha$-$S_{8}$
crystals and compared it with that of monoclinic $\alpha$'-$S_{8}$
\cite{pap1,pap2,pap3,pap4}, a previously proposed phase that is calculated
with lower mean volume per molecule and lower potential energy than
$\alpha$-$S_{8}$ crystals. The monoclinic $\alpha$'-$S_{8}$ crystal
was obtained in the course of MD simulations, using simple inter-
and intramolecular potential models, but has not been experimentally
observed \cite{alfa2,exp-alemanes}. Therefore, in this work we performed
an accurate measurement of the free energy difference between both
crystals at 100 and 300 K (above and below the spontaneous transition
found in ref. \cite{pap1}), as given by the simple molecular model
of \cite{pap1}. The thermodynamic integration method \cite{free-frenkel-ladd1,free-frenkel-correction}
is implemented, taking as reference system the corresponding Einstein
crystals. For the sake of comparison we also included calculations
of the free energy in the quasi-harmonic approximation, a fast estimation
method that can be easily implemented in MD simulations.

The accurate free energy calculations performed here still show that
monoclinic $\alpha$'-$S_{8}$ is calculated more stable than the
orthorhombic experimentally observed $\alpha$-$S_{8}$, in accord
with our previous calculations \cite{pap1,pap2} performed with the
same simple molecular model and in the quasi-harmonic approximation. 

As regards the accuracy of the free energy calculations, we found
that for the thermodynamic integration method the results are extremely
sensitive to the accurate calculation of the values $U(\lambda)$.
In particular for $\lambda\sim1$ long MD simulations are needed.
Although the absolute free energy values differ for both calculation
methods, both measure a difference of about 2 kJ/mol between $\alpha$-
and $\alpha$'-$S_{8}$ crystals, being $\alpha$'-$S_{8}$ the more
stable.

The approximated method for calculating the free energy \emph{via}
the contribution of the vibrational modes to the energy and entropy
of the system turns out to be very useful to determine relative values
between polymorphic phases or between different temperatures of a
given sample. This result is very important, taking into account that
this method is, at least, ten times faster than the thermodynamic
integration method, that requires at least ten lengthly MD simulations
(one MD run for each value of the $\lambda$ parameter), in order
to calculate the free energy of one sample at a given temperature.

\begin{acknowledgments}
C. P. thanks Ignacio Urrutia for fruitful and encouraging discussions
as regards bond constrains in free energy calculations and FOSDIC
for a partial support. This work was partially supported by the grant
PIP 0859/98 of CONICET and Fundación Balseiro-2001.

\appendix
\end{acknowledgments}

\section{APPENDIX\label{sec:APPENDIX}}

We estimate here the difference in the calculated statistical averages
for MD simulations performed with and without molecular bond length
constraints. This topic has been extensively treated in literature
and is not of minor importance \cite{fe-const2,fe-const3,fe-const4,understanding}.
Indeed, it has been widely studied in the case of free energy calculations
that involve a reaction coordinate \cite{fe-const3,fe-const4}.

We recall the results of chapter 10 from ref. \cite{understanding}
and the appendix from ref. \cite{fe-const2}. The expression for the
ratio of molecular position probability for constrained and unconstrained
system is:

\begin{equation}
\frac{\rho(\mathbf{q}_{1},...,\mathbf{q}_{3N})}{\rho(\mathbf{q}_{1},...,\mathbf{q}_{3N-l,}\mathbf{q}_{1}^{c}=\sigma_{1},...,\mathbf{q}_{l}^{c}=\sigma_{l})}=c\,\sqrt{\left|Z\right|}\,,\label{eq:ap1}\end{equation}

where we explicitly note the constrained degrees of freedom with $\mathbf{q}_{i}^{c}=\sigma_{i}$,
being the standard bond length constraint,

\begin{equation}
\sigma_{i}=\sqrt{m_{i}}(\sqrt{\mathbf{r}_{ij}^{2}}-d\equiv0)\,,\label{eq:ap-sigma}\end{equation}

and $l$ constrained bonds are assumed. The $l\times l$ $Z$ matrix,
in the right hand side of \ref{eq:ap1} is defined as \cite{understanding,fe-const2}:

\begin{equation}
Z_{mn}=\sum_{i=1}^{N}\frac{1}{m_{i}}(\frac{\partial\sigma_{m}}{\partial\mathbf{r}_{i}}).(\frac{\partial\sigma_{n}}{\partial\mathbf{r}_{i}})\:,\label{eq:ap-z-def}\end{equation}
and $\left|\right|$~stands for the determinant. The origin of $Z$
is from the fact that not only $\sigma(\mathbf{r}_{i})=0$ must be
required to constraint the bonds in the simulations, but also $\dot{\sigma}(\mathbf{r}_{i})=0$
at all times \cite{fe-const4}. It should be noted that $\mathbf{q}$
in eq. \ref{eq:ap1} is used for generalized coordinates while $\mathbf{r}_{i}$
for cartesian ones.

For the $S_{8}$ molecule we have 24 degrees of freedom, from which
8 have been constrained in our model (see sec. \ref{sec:Intra--and-intermolecular}).
Therefore we have 16 degrees of freedom with 8 bond constraints following
eq. \ref{eq:ap-sigma}, with $d=$2.06~\AA. Matrix $Z$ can be explicitly
calculated for a $S_{8}$ molecule from eq. \ref{eq:ap-z-def}:

\[
\left(\begin{array}{cccccccc}
2 & \cos(\beta_{12}) & 0 & 0 & 0 & 0 & 0 & \cos(\beta_{18})\\
\cos(\beta_{12}) & 2 & \cos(\beta_{23}) & 0 & 0 & 0 & 0 & 0\\
0 & \cos(\beta_{23}) & 2 & \cos(\beta_{34}) & 0 & 0 & 0 & 0\\
0 & 0 & \cos(\beta_{34}) & 2 & \cos(\beta_{45}) & 0 & 0 & 0\\
0 & 0 & 0 & \cos(\beta_{45}) & 2 & \cos(\beta_{56}) & 0 & 0\\
0 & 0 & 0 & 0 & \cos(\beta_{56}) & 2 & \cos(\beta_{67}) & 0\\
0 & 0 & 0 & 0 & 0 & \cos(\beta_{67}) & 2 & \cos(\beta_{78})\\
\cos(\beta_{18}) & 0 & 0 & 0 & 0 & 0 & \cos(\beta_{78}) & 2\end{array}\right),\]

where $\beta_{ij}$ is the bending bond angle between consecutive
angles $i,\, j$ and $j+1$.

If we take a mean $\beta$ value for all the bending angles and evaluate
the determinant in eq. \ref{eq:ap1}, we get:

\begin{equation}
\left|Z\right|=256-576\cos(\beta)^{2}+400\cos(\beta)^{4}-88\cos(\beta)^{6}\,.\label{eq:det-z}\end{equation}

We can now take a statistical mean value over the NVT ensemble, averaging
eq. \ref{eq:det-z} over a MD run and normalizing with the maximum
value. This provides an estimation of the effect of constraints over
statistical averages such as used in eq. \ref{eq:ti4}, to carry out
the thermodynamic integration. We got the following main difference
between the constrained and unconstrained probabilities at $T=300K$:

\[
\frac{\rho}{\rho_{constrained}}\sim\left\langle \sqrt{\left|Z\right|}\right\rangle \sim0.954\,.\]

The closeness to 1 indicates that the effect of bond constrains is
practically negligible for simulations in the studied range of thermodynamic
variables. From this estimation we can conclude that the points obtained
in figure \ref{fig-u-free} are almost identical to those that can
be obtained in a similar simulation without bond constrains, that
would have been quite more expensive in CPU time. This simple calculation,
that we present here for close ring-molecules, is complementary to
that already obtained in ref. \cite{const-fixman} for open chain
molecules.

\subsection*{Contribution of stretching modes to free energy}

We include here an estimation of the free energy difference between
samples with and without molecular bond constrains, under the framework
of quasi-harmonic approximation commented in sec. \ref{sub:Quasi-harmonic-approximation}.
If we think in the normal modes as harmonic oscillators of frequency
$\nu_{i}$, we have a {}`trivial' partition function $Z_{i}$ for
each oscillator of the form\cite{bornhuang}:

\[
Z_{i}=e^{-i\frac{1}{2}h\nu_{i}}\sum_{s=0}^{\infty}e^{-sh\nu_{i}/kT}=\frac{e^{-i\frac{1}{2}h\nu_{i}}}{1-e^{-ih\nu_{i}}}\]

where $h$ is the Planck constant, $k$ the Boltzmann factor and $s$
the positive integer quantum number of the oscillator of frequency
$\nu_{i}$. For a system of such independent oscillators we have a
partition function:

\begin{equation}
Q_{simul}=\int_{0}^{\infty}D(\nu)(\frac{e^{-\frac{1}{2}\beta h\nu}}{1-e^{-\beta h\nu}})\, d\nu\,,\label{eq:q-harm}\end{equation}

where $D(\nu)$ is the density of vibrational states already defined
in sec. \ref{sub:Quasi-harmonic-approximation}and $Q_{simul}$ is
calculated from our bond constrained MD simulation. The partition
function of the unconstrained system $Q$ can be estimated in the
following way: taking into account that for the $S_{8}$ molecule
the stretching vibrational modes are well above bending and torsional
modes in the $\nu$ scale (see sec. \ref{sec:Intra--and-intermolecular}),
we have separable contributions for the total partition function of
the molecule and

\begin{equation}
Q=\int_{0}^{\nu_{max}}D_{simul}(\nu)(\frac{e^{-\frac{1}{2}\beta h\nu}}{1-e^{-\beta h\nu}})\, d\nu+\int_{_{\nu_{max}}}^{\infty}D_{stretch}(\nu)(\frac{e^{-\frac{1}{2}\beta h\nu}}{1-e^{-\beta h\nu}})\, d\nu\equiv Q_{simul}+Q_{stretch},\,\label{eq:q-tot-harm}\end{equation}

where we have assumed that the interval $(0,\nu_{max})$ accounts
for the bending, torsional (dihedral) and lattice vibration modes
and the high frequencies of bond stretching are {}`isolated' in
the interval $(\nu_{max},\infty)$. The first term in eq. \ref{eq:q-tot-harm}
is obtained from our simulation and the second one can be simply approximated
by:

\[
Q_{stretch}=N^{stretch}(\frac{e^{-\frac{1}{2}\beta h\nu_{stretch}}}{1-e^{-\beta h\nu_{stretch}}})\,,\]

here $N^{stretch}$ are the 8 stretching degrees of freedom (constrained
in the MD simulations) and $\nu_{stretch}$ is the mean stretching
frequency. This is equivalent to use the relationship $D_{stretch}(\nu)=N^{stretch}\times\delta(\nu-\nu_{stretch})$.
The free energy difference between considering or not the stretching
degrees of freedom can be estimated, under the quasi-harmonic approximation
by:

\[
\beta\Delta F=\beta(F_{tot}-F_{simul})=\ln(\frac{Q_{simul}}{Q_{simul}+Q_{stretch}})\]

For $T=300K$ and the sulphur mean stretching frequency $\nu_{stretch}=453.8\, cm^{-1}$,
we get $\beta\Delta F=0.075$ which is a relatively and also an absolute
small number when compared with the $\beta F$ values of table \ref{tab-detalle-ti}.

This simple calculation can be considered complementary to those of
ref. \cite{constraints-estima1,constraints-estima3,constraint-estima2},
that provided a thorough discussion of the role of bond constrains
in free energy calculations.

\end{document}